\documentclass[11pt,a4paper]{article}

\usepackage{amsmath,amssymb,tikz,hyperref,empheq,mathrsfs,amsthm}
\usepackage{graphicx}
 \allowdisplaybreaks
\def\be{\begin{equation}}
\def\ee{\end{equation}}
\def\ba#1\ea{\begin{align}#1\end{align}}
\def\bg#1\eg{\begin{gather}#1\end{gather}}
\def\bm#1\em{\begin{multline}#1\end{multline}}
\def\bmd#1\emd{\begin{multlined}#1\end{multlined}}

\def\({\left(}
\def\){\right)}
\def\[{\left[}
\def\]{\right]}

\def\Tr{{\rm Tr}}

\def \be {\begin{equation}}
\def \ee {\end{equation}}
\def \ba {\begin{array}}
\def \ea {\end{array}}
\def \bea{\begin{eqnarray}}
\def \eea{\end{eqnarray}}

\def\bea{\begin{eqnarray}}
\def\eea{\end{eqnarray}}

\newcommand{\bit}{\begin{itemize}}  \newcommand{\eit}{\end{itemize}}
\newcommand{\ben}{\begin{enumerate}}  \newcommand{\een}{\end{enumerate}}

\long\def\symbolfootnote[#1]#2{\begingroup%
\def\thefootnote{\fnsymbol{footnote}}\footnote[#1]{#2}\endgroup}

\newcommand{\zju}{{$^a$ \it Department of Physics, Zhejiang University, Hangzhou 310027, P. R. China}}

\newcommand{\wu}{{$^b$\it School of Science, Westlake University, Hangzhou 310024, P. R. China}}

\newcommand{\wias}{{$^c$\it Institute of Natural Sciences, Westlake Institute of Advanced Study, Hangzhou 310024, \\ P. R. China}}

\newcommand{\email}{\emph{E-mails}: tangyin@westlake.edu.cn, tangqicheng@westlake.edu.cn, \\ zhuwei@westlake.edu.cn}

\begin{document}
\thispagestyle{empty}
\begin{center}

~\vspace{20pt}

{\Large\bf Thermal Correction to Entanglement Spectrum for Conformal Field Theories }

\vspace{25pt}

Yin Tang$^{a,b,c}$ , 
Qicheng Tang$^{a,b,c}$ 
and Wei Zhu$^{b,c}$

\vspace{10pt}${}$\zju
\\
\vspace{10pt}${}$\wu
\\
\vspace{10pt}${}$\wias
\\
\vspace{10pt}${}$\email

\vspace{3mm}

\begin{abstract}

We calculate the thermal correction to the entanglement spectrum for separating a single interval of two dimensional conformal field theories. 
	Our derivation is a direct extension of  the thermal correction to the R\'enyi entropy. 
	Within a low-temperature expansion by including only the first excited state in the thermal density matrix, we approach analytical results of the thermal correction to the entanglement spectrum at both of the small and large interval limit. 
	We find the temperature correction reduces the large eigenvalues in the entanglement spectrum while increases the small eigenvalues in the entanglement spectrum, leading to an overall crossover changing pattern of the entanglement spectrum. Crucially, at low-temperature limit, the thermal corrections are dominated by the first excited state and depend on its scaling dimension $\Delta$ and degeneracy $g$. This opens an avenue to extract universal information of underlying conformal data via the thermal entanglement  spectrum.
	All of these analytical computation is supported from numerical simulations using 1+1 dimensional free fermion. 
	Finally, we extend our calculation to resolve the thermal correction to the symmetry-resolved entanglement spectrum.
\end{abstract}

\end{center}

\newpage
\setcounter{footnote}{0}
\setcounter{page}{1}

\tableofcontents

	\section{Introduction}
\label{sec:intro}

Quantum entanglement, an inherently quantum mechanical nonlocal phenomenon, plays a central role in modern physics. 
Besides the general interests in quantum information theory \cite{nielsen_chuang_2010,Horodecki:2009zz}, nowadays the concepts and anaysis of entanglement has been extended to cover diverse topics, ranging from describing collective behaviors in condensed matters \cite{Amico:2007ag, Laflorencie:2015eck} to building up spacetime of the universe \cite{Ryu:2006bv,Ryu:2006ef,Hubeny:2007xt,Nishioka:2009un,Rangamani:2016dms,Nishioka:2018khk}. 
In particular, the study of quantum entanglement brings novel insights into understanding critical phenomena in many-body systems. 
Especially for $(1+1)$-dimensional [$(1+1)$D] critical systems in equilibration \cite{Calabrese:2004eu,Calabrese:2005zw,Calabrese:2009qy,Vidal:2002rm,Fradkin:2006mb,Hsu:2008af} and out-of-equilibrium \cite{Calabrese:2005in,Calabrese:2007rg,Eisler:2007,Cardy:2014rqa,Cardy:2015xaa,Cardy:2016fqc,Calabrese:2016xau,Wen:2018svb,Tang:2019nri}, the intrinsic conformal invariance leads to rigid constrains on physical proprieties, including the structure of quantum entanglement. 
One of the most important measures of quantum entanglement is the \textit{entanglement entropy} (EE), defined as the von Neumann entropy of the reduced density matrix of the subsystem $\rho_A$: 
\begin{equation}
S = - \Tr \rho_A \ln \rho_A. 
\end{equation}
It is analytically predicted that the scaling behavior of the EE in a $(1+1)$D critical ground state is fully determinated by the central charge $c$ of the underlying conformal field theory (CFT) as 
$ S = \frac{c}{3} \ln l_{\rm eff} $ \cite{Holzhey:1994we,Calabrese:2004eu,Calabrese:2005zw,Calabrese:2009qy,Casini:2009sr},
where $l_{\rm eff}$ is the effective length of the subsystem. 
This result leads to great achievement on understanding quantum criticality via entanglement, especially for minimal models that are fully characterized by a single number of the central charge $c$.

Besides the EE, there are also other measures of quantum entanglement that reflects interesting critical phenomena. 
Specifically, the \textit{entanglement spectrum} (ES) has attracted considerable attentions. 
The ES denotes the eigenvalues of the \textit{entanglement Hamiltonian} $H_E$ that is defined by $\rho_A = e^{-H_E}$. 
Obviously, the EE is just a kind of statistical average of the ES, so that it is naturally to expect that the ES should contain more information \cite{Li:2008kda,Fidkowski:2010,Roy:2020frd,Pollmann:2010,Yao:2010,regnault2015entanglement}. 
In this context, it is interesting to consider whether the ES -- as a more precise measure of quantum entanglement than the EE -- can be determinated by using the conformal symmetry. 
In $(1+1)$D critical systems, it is found that the ES can be indeed solved through CFT techniques \cite{Wen:2018svb}, and the conformal data can be resolved by the ES \cite{Tang:2019nri}. 
Moreover, at the continuum limit, the gap of entanglement spectrum closes and the continuous distribution of the ES is found to be dependent only on the central charge \cite{Calabrese:2008}. 
These results demonstrate the powerful constrain of conformal invariance for $(1+1)$D critical systems.

%
The recent progress on quantum simulations allows to realize critical systems experimentally. 
These advanced techniques, such as optical lattices~\cite{Bloch_2008_review_optical_lattice, Zoller2012quench_EE_optical_lattice}, superconducting circuits~\cite{Google2019_quantum_supremacy}, and trapped ions~\cite{Lukin_2017_51bit}, are expected to provide a direct access to the entanglement content of many-body states. 
Although these systems are prepared at extremely low temperature, the thermal corrections are still necessary to be addressed for probing the emergent criticality. 
In particular, at finite temperature, the ideal pure state becomes a thermal density matrix. The question of whether the observation in ideal (critical) models at zero temperature are stable under temporal perturbation remains open.

In this regard, in the present work we aim to address this question by studying the ES of $(1+1)$D critical systems at finite temperatures. 
First of all, by using the conformal symmetry, we analytically derive the thermal correction to the ES by cutting a small region in a finite system.  
Second, these predictions are verified by the numerical simulations at a low temperature $T = \beta^{-1} \sim 1/L$ comparing to the energy gap for a finite total system. 
These results are not only expected to be valuable for understanding the thermal effect theoretically, but also potentially useful for detecting quantum criticality experimentally.

\section{Preliminary: some known results on the entanglement in $(1+1)$D CFT}

Historically, the attempt to understand quantum entanglement in many-body systems started with the calculation of the EE. 
By definition, it requires the knowledge of the full spectrum of the reduced density matrix. 
However, a direct determination of the spectrum of the reduced density matrix is generally hard, which motivates an application of the \textit{replica trick} to study quantum entanglement in field theory \cite{Holzhey:1994we,Calabrese:2004eu,Calabrese:2005zw}. 
It is important to observe that the EE can be treated as a special case of the R\'enyi entropy $S_n$ by considering 
\begin{equation}\label{eq:replica}
S = \lim_{n \to 1} \frac{\partial}{\partial n} \Tr \rho_A^n
= \lim_{n \to 1} \frac{\ln \Tr \rho_A^n }{1 - n} 
= \lim_{n \to 1} S_n . 
\end{equation}
For positive integer $n$, the value of $\Tr \rho_A^n$ can be computed by making $n$ copies (replica) of the theory and sewing them along the entanglement cut along the boundary of $A$. The analytical continuation of $n$ is then assumed for applying the derivative on it, with taking the replica limit $n \to 1$. 
Specifically, making $n$ copies of the theory actually forms a non-trivial $n$-fold manifold, so called \textit{replica manifold}.  
To calculate $\Tr \rho_A^n$ for ground state wavefunction, we further consider an Euclidean path integral representation. It gives
\begin{equation}
\Tr \rho_A^n = \frac{Z_n}{(Z_1)^n} ,
\end{equation}
where $Z_n$ is the partition function on the $n$-fold replica manifold, and $Z_1$ is for a single sheet. 
The determination of $Z_n$ is generally hard, however, for two-dimensional conformal field theories it can be calculated by using a conformal mapping of correlation functions from the usual (Euclidean) flat spacetime to the replica manifold. It gives
\begin{equation}\label{eq:zero_tem_trace_rho}
\Tr \rho_A^n = \sum_{i} \lambda_i^n = 
c_n l_{\text{eff}}^{\frac{c(1-n^2)}{6n}} , 
\end{equation}
where $c$ is the central charge, $c_n$ is a non-universal constant that will be ignored in our further consideration, and the effective length of the subsystem $l_{\text{eff}}$ is determinated by the geometry of the manifold with entanglement cut. 
For example, when cutting a single finite region with length $l$ from an infinite circle (with periodic boundary condition), we have $l_{\text{eff}} = \frac{l}{a}$ with the lattice constant $a$ (In the rest of this paper, we will simply choice $a = 1$ as the length scale). 
Combine with the Eq.~\eqref{eq:replica}, we have 
\begin{equation}
S = \lim_{n \to 1} \frac{\partial}{\partial n} 
\left[ c_n l_{\text{eff}}^{\frac{c(1-n^2)}{6n}} \right] \\ 
\sim \frac{c}{3} \ln l_{\text{eff}} . 
\end{equation}
Meanwhile, if we take $n \to \infty$ in Eq.~\eqref{eq:replica}, it gives the maximum of eigenvalues 
\begin{equation}\label{eq:lambda_max}
- \ln \lambda_1
= \lim_{n \to \infty} S_n 
= \lim_{n \to \infty} \frac{\ln \Tr \rho_A^n}{1-n} 
= \ln l_{\text{eff}}^{\frac{c}{6}} 
\quad \Rightarrow \quad 
\lambda_1 = l_{\text{eff}}^{-\frac{c}{6}} .
\end{equation}
This leads to the result of $S = -2\ln \lambda_1$ that is observed in critical spin chains.

Although the original proposal of introducing the replica trick is to avoid the calculation of ES, the determination of $\Tr \rho_A^n$ is actually applicable to give the ES of $2$D CFTs \cite{Calabrese:2008}. 
For accessing the spectrum information of the reduced density matrix, we investigate the distribution function $P(\lambda) =\sum_i \delta(\lambda - \lambda_i)$ for the eigenvalues $\{ \lambda_i \}$. 
The trick on evaluating the distribution $P(\lambda)$ is to consider the following construction 
\begin{equation}\label{eq:spectrum_distribution}
z P(z) 
= \frac{1}{\pi} \lim_{\epsilon \to 0} 
\Im \int d\lambda \frac{\lambda P(\lambda)}{z - \lambda} 
= \frac{1}{\pi} \lim_{\epsilon \to 0} 
\Im \sum_{n=1}^\infty R_n  z^{-n} , 
\end{equation}
with letting $z = \lambda - i\epsilon$. Here the Laurent expansion coefficient is equal to the $n$-th momentum of the reduced density matrix $\rho_A$
\begin{equation}
R_n = \int d\lambda [\lambda^n P(\lambda)] = \Tr \rho_A^n . 
\end{equation}
This means that knowing $\Tr \rho_A^n$ for general $n$ can fully determinate the eigenvalues $\{\lambda_i\}$ of the reduced density matrix, which is equivalent to solving the ES: $\xi_i = -\ln \lambda_i$. 
After some algebra, it gives
\begin{equation}
\label{eq:es_0K}
P(\lambda) = \delta(\lambda_1 - \lambda) 
+ \theta(\lambda_1 - \lambda) \frac{- \ln \lambda_1}
{ \lambda \sqrt{ - \ln \lambda_1 \ln \frac{\lambda_1}{\lambda} } } 
I_1\left[ 2 \sqrt{ - \ln \lambda_1 \ln \frac{\lambda_1}{\lambda} } \right] ,
\end{equation}
where $I$ is the modified Bessel function of the first kind. 
It is obvious that both the value of EE and the distribution of ES are only dependent by the largest eigenvalues of the reduced density matrix, which is given by the central charge of underlying CFT: $\lambda_1 = l_{\text{eff}}^{-\frac{c}{6}}$.

The thermal correction to the Renyi entropy for a gapped theory was firstly considered by Herzog and Spillane in \cite{Herzog:2012bw}, where they conjectured the correction scales as $e^{-\beta m}$ and provided numerical evidence for a massive scalar in 1+1 dimensions. Subsequent research conformed to this point both from field theory \cite{Herzog:2013py,Cardy:2014jwa,Datta:2013hba,Chen:2014unl,Herzog:2014tfa,Herzog:2014fra,Chen:2014hta,Herzog:2015cxa,Chen:2016lbu} and holography \cite{Azeyanagi:2007bj,Barrella:2013wja,Datta:2013hba,Chen:2014unl,Chen:2016lbu,Klich:2017qmt,Chen:2017ahf}. In \cite{Cardy:2014jwa}, Cardy and Herzog considered the following low-temperature expansion 
\begin{equation}
\rho = \frac{e^{-\beta H_{\text{CFT}}}}{\Tr e^{-\beta H_{\text{CFT}}}} 
= \frac{|0 \rangle \langle 0| 
	+ |\psi \rangle \langle \psi| e^{-2\pi \Delta \beta/L} + \cdots}
{1 + e^{-2\pi \Delta \beta/L} + \cdots} ,
\end{equation}
where $|\psi\rangle$ is the first excited state that denoted by the conformal wight $\Delta$, and the total system is chosen to have a finite size $L$ to form a finite gap $m = 2\pi \Delta /L$ from the ground state $|0\rangle$. To the lowest order, it leads to the thermal correction of $\Tr \rho_A^n$ as
\begin{equation}
\begin{aligned}
& \quad \Tr \rho_A^n (\beta)  = \frac{\Tr \left[ \Tr_B \left( 
	|0 \rangle \langle 0| 
	+ g |\psi \rangle \langle \psi| e^{-2\pi \Delta \beta/L} + \cdots
	\right) \right]^n}
{\left( 1 +g e^{-2\pi \Delta \beta / L} + \cdots \right)^n} \\
& = \Tr \left( \Tr_B |0 \rangle \langle 0| \right)^n 
\left[ 1 + g \left( \frac{\Tr \left[ \Tr_B |\psi \rangle \langle \psi| \left( \Tr_B |0 \rangle \langle 0| \right)^{n-1} \right]}{\Tr \left( \Tr_B |0 \rangle \langle 0| \right)} - 1 \right) n e^{-2\pi \Delta \beta/L} + \cdots \right] , 
\end{aligned}
\end{equation}
where $g$ is the degeneracy of first excited state (the ground state is considered to be unique). 
Here the leading term is the zero temperature contribution of Eq.~\eqref{eq:zero_tem_trace_rho} for cutting a single finite region with length $l$ from a ring with circumference $L$
\begin{equation}
\Tr \left( \Tr_B |0 \rangle \langle 0| \right)^n  = \Tr \rho_A^n(\beta=0) = \sum_i \lambda_i^n =   l_{\text{eff}}^{\frac{c(1-n^2)}{6n}} = (\frac{L}{\pi}\sin \frac{\pi l}{L})^{\frac{c(1-n^2)}{6n}} .
\end{equation}
Similar to the dominate contribution, here the subleading finite temperature correction is equivalent to a correlation function on a $n$-fold replica manifold that can be calculated by a conformal mapping.  The result is 
\begin{equation}
\frac{\Tr \left[ \Tr_B |\psi \rangle \langle \psi| \left( \Tr_B |0 \rangle \langle 0| \right)^{n-1} \right]}{\Tr \left( \Tr_B |0 \rangle \langle 0| \right)} 
= \frac{1}{n^{2\Delta}} \frac{\sin^{2\Delta}{\frac{\pi l}{L}}}{\sin^{2\Delta}{\frac{\pi l}{nL}}} ,
\end{equation}

Next, we will show this thermal expansion, combining with the Eq.~\eqref{eq:spectrum_distribution}, leads to the thermal correction of the ES at low-temperature.

\section{The entanglement spectrum in thermal states at low-temperature}

In this section, we will provide an analytical derivation of the ES for thermal critical states. 
As discussed in the previous section, at low-temperature we have
\begin{equation}\label{eq:thermal_rho_n}
\Tr \rho_A^n(\beta) = 
\left( \frac{L}{\pi}\sin \frac{\pi l}{L} \right)^{\frac{c(1-n^2)}{6n}}
\left[ 1+g \left( 
\frac{1}{n^{2\Delta}} \frac{\sin^{2\Delta}{\frac{\pi l}{L}}}{\sin^{2\Delta}{\frac{\pi l}{nL}}}-1 
\right)
ne^{-2\pi \Delta \beta /L}  
\right] + \cdots . 
\end{equation}
The first term in Eq.~\eqref{eq:thermal_rho_n} is the zero-temperature result, and the latter terms come from the finite temperature $\beta$. 
Here, we will use a similar scheme of calculating the zero-temperature ES \cite{Calabrese:2008} from $\Tr \rho_A^n$ to evaluate the thermal correction to ES.

Since the zero-temperature ES is known in previous investigations, here we will only focus on the thermal correction. 
In the lowest-order expansion, the Eq.~\eqref{eq:thermal_rho_n} can be rewritten to 
\begin{equation}
\Tr \rho_A^n(\beta) 
= \sum_i \left[ \lambda_i + \delta \lambda_i(\beta) \right]^n 
\approx \sum_i \lambda_i^n \left[ 1 + \frac{n \delta \lambda_i(\beta)}{\lambda_i} \right] 
= \sum_i \lambda_i^n \left[ 1 + 
\eta_i  n e^{-2\pi \Delta \beta/ L}
\right] ,
\end{equation}
with a temperature-independent parameter $\eta_i = \frac{\delta \lambda_i(\beta)}{\lambda_i} e^{2\pi \Delta \beta/L}$ that satisfies
\begin{equation}\label{eq:lambda_eta_0}
\sum_i \lambda_i^n \eta_i 
= \left( \frac{L}{\pi}\sin \frac{\pi l}{L} \right)^{\frac{c(1-n^2)}{6n}} 
g \left( 
\frac{1}{n^{2\Delta}} \frac{\sin^{2\Delta}{\frac{\pi l}{L}}}{\sin^{2\Delta}{\frac{\pi l}{nL}}}-1 
\right) .
\end{equation}
This thermal correction does not change the normalization condition of the reduced density matrix $\Tr \rho_A(\beta) = 1$, since we have $\lim_{n \to 1} \sum_i \lambda_i^n \eta_i = 0$. 
Moreover, similar to calculating the largest eigenvalue $\lambda_1$ in Eq.~\eqref{eq:lambda_max}, here we have 
\begin{equation}\label{eq:eta_1}
\eta_1 =  \lim_{n \to \infty} 
g \left( 
\frac{1}{n^{2\Delta}} \frac{\sin^{2\Delta}{\frac{\pi l}{L}}}{\sin^{2\Delta}{\frac{\pi l}{nL}}}-1 
\right) 
= g \left[ 
\frac{\sin^{2\Delta}  \left( \pi l/L \right) }
{ \left( \pi l/L \right)^{2\Delta} } 
- 1 \right] . 
\end{equation}
The corrected largest eigenvalue is immediately obtained
\begin{equation}
\lambda_1(\beta) \approx 
\lambda_1 \left( 1 + \eta_1 e^{-2\pi \Delta \beta/L} \right)
=  \left( \frac{L}{\pi}\sin \frac{\pi l}{L} \right)^{-\frac{c}{6}} 
\left\{ 1 + g \left[ 
\frac{\sin^{2\Delta}  \left( \pi l/L \right) }
{ \left( \pi l/L \right)^{2\Delta} } 
- 1 \right] e^{-2\pi \Delta \beta/L} \right\} .
\end{equation}

As a rough estimation, one can consider that the EE of thermal states is (approximately) determinated by the largest eigenvalue $\lambda_1(\beta)$ as the ground state
\begin{equation}
-\ln \lambda_1(\beta) 
\approx -\ln \lambda_1 - \ln (1 + \eta_1 e^{-2\pi \Delta \beta/L}) 
\approx -\ln \lambda_1 - \eta_1 e^{-2\pi \Delta \beta/L} . 
\end{equation}
It then follows 
\begin{equation}\label{eq:rough_EE_beta}
\delta S(\beta) \approx - \eta_1 e^{-2\pi \Delta \beta/L} 
= g \left[ 
\frac{\sin^{2\Delta}  \left( \pi l/L \right) }
{ \left( \pi l/L \right)^{2\Delta} } 
- 1 \right] e^{-2\pi \Delta \beta/L} . 
\end{equation}
At the lowest-order of a small interval limit $l \ll L$, we find that this is in line with the previously known exact solution~\cite{Cardy:2014jwa}
\begin{equation}\label{eq:exact_EE_beta}
\delta S(\beta) = 2 g \Delta \left[ 1 - \frac{\pi l}{L} \cot \left( \frac{\pi l}{L} \right) \right] 
e^{-2\pi \Delta \beta/L} + \mathcal{O}( e^{-2\pi \Delta \beta/L} ) . 
\end{equation}
However, for higher-order expansions of $\mathcal{O}\left( \frac{l}{L} \right)^4$, Eq.~\eqref{eq:rough_EE_beta} and \eqref{eq:exact_EE_beta} are not consistent with each other. 
Physically, this is because that the entanglement content of a thermal state contains more information than the pure ground state, and cannot fully determinated by a single number.

Below we try to solve the thermal correction to the entanglement spectrum for each level $i$. Here, we consider the distribution function $Q(\lambda) = \sum_i \delta(\lambda - \lambda_i) \eta_i$, which is expected to be temperature-independent at the lowest-order of $e^{-2\pi \Delta \beta/L}$. 
Analog to Eq.~\eqref{eq:spectrum_distribution}, we consider the following construction 
\begin{equation}
\lambda Q(\lambda) 
= \lim_{\epsilon \to 0} \Im f(z=\lambda - i\epsilon)
\end{equation}
\begin{equation}
f(z) = \frac{1}{\pi} \lim_{\epsilon \to 0}  \int d\lambda \frac{\lambda Q(\lambda)}{z - \lambda} 
= \frac{1}{\pi} \lim_{\epsilon \to 0} \sum_{n=1}^\infty \left( \sum_i \lambda_i^n \eta_i \right) z^{-n}
\end{equation}
Although each term in the above expansion is known, it is not easy to obtain a compact form with the sine functions in $l_{\text{eff}}$.  
Hence, we will focus on the small or large interval limits, where the effective length can be simplified to power functions.

\subsection{Small interval limit}

At the \textit{small interval limit} $l \ll L$, we have 
\begin{equation}
\sum_i \lambda_i^n \eta_i 
= g 
\frac{\pi^2(1-n^2)\Delta}{3n^2} 
l^{\frac{c(1-n^2)}{6n}} 
\left( \frac{l}{L} \right)^2 
+ \mathcal{O} \left( \frac{l}{L} \right)^4 .
\end{equation}
Use the same trick as in the zero-temperature case, this leads to
\begin{equation}
\begin{aligned}
\lambda Q(\lambda) 
& \approx  \frac{1}{\pi} \lim_{\epsilon \to 0} \Im \sum_{n=1}^\infty 
g 
\frac{\pi^2(1-n^2)\Delta}{3n^2} 
l^{\frac{c(1-n^2)}{6n}} 
\left( \frac{l}{L} \right)^2 
(\lambda - i\epsilon)^{-n} \\ 
& =  
\frac{\pi g \Delta}{3} \left( \frac{l}{L} \right)^2 
\lim_{\epsilon \to 0} \Im 
\left\{ 
\sum_{k=0}^{\infty} 
\frac{\left( - \ln \lambda_1 \right)^k}{k\mathrm{!}} 
\left[
\mathrm{Li}_{k+2} \left( \frac{\lambda_1}{\lambda-i\epsilon} \right) 
- \mathrm{Li}_{k} \left( \frac{\lambda_1}{\lambda-i\epsilon} \right)
\right] 
\right\} \\
& = 
\frac{\pi^2 g \Delta}{3} \left( \frac{l}{L} \right)^2  
\left[
- \lambda \delta(\lambda_1 - \lambda) + 
\theta(\lambda_1 - \lambda) 
I_1\left( 2 u \right)
\left( 
\frac{u}{-\ln\lambda_1} 
- \frac{-\ln\lambda_1}{u}
\right)
\right] ,
\end{aligned}
\end{equation}
where $u = \sqrt{-\ln \lambda_1 \ln \frac{\lambda_1}{\lambda}}$. 
A direct comparison of the distribution function to numerical results on discrete lattice model is hard, instead, we consider the integral of 

\begin{equation}
s_{\eta}(M) = \int_{\lambda_M}^{\lambda_1} Q(\lambda) d\lambda
= \frac{\pi^2 g \Delta}{12 \left( -\ln \lambda_1 \right)^2} 
\left( \frac{l}{L} \right)^2 
\left\{
[ I_0^{-1}(M) ]^2  I_2[ I_0^{-1}(M) ] 
- 4 \left( -\ln \lambda_1 \right)^2 M \right\} .
\end{equation}
It then follows to give the iterative solution of $\eta_i$ and $\lambda_i$
\begin{equation}
\label{eq:eta_i}
\eta_i = \frac{\pi^2 g\Delta}{12\left(-\ln\lambda_1\right)^2} 
\left(\frac{l}{L}\right)^2 
\left\{ 
[I_0^{-1}(i)]^2 I_2 [I^{-1}_0(i)] 
-[I_0^{-1}(i-1)]^2 I_2[I^{-1}_0(i-1)] 
-4\left( -\ln \lambda_1 \right)^2 
\right\} .
\end{equation}
\begin{equation}
\label{eq:lambda_i}
\lambda_i(\beta) \approx \lambda_i (1+\eta_i e^{-2\pi \Delta \beta / L})
\end{equation}
By using $I_0(0) = 1$, $I_m(0)=0 (m = 1, 2, \dots)$, it is easy to get 
\begin{equation}
\eta_1 = s_{\eta}(1) 
\approx - \frac{\pi^2 g \Delta}{3} 
\left( \frac{l}{L} \right)^2 \ ,
\end{equation}
which is consistent with the result in Eq.~\eqref{eq:eta_1} at the small interval limit $l \ll L$.

\subsection{Large interval limit}

Since for thermal state the symmetry of $S(l) = S(L-l)$ is broken by mixing the excited states, it is also important to investigate the ES at \textit{large interval limit} $l \to L$. 
The expansion of Eq.~\eqref{eq:lambda_eta_0} (for $n \geq 2$) in terms of $\frac{L-l}{L}$ is given by 
\begin{equation}
\sum_i \lambda_i ^n\cdot \eta_i
= \left( \frac{L}{\pi a}\sin \frac{\pi l}{L} \right)^{\frac{c(1-n^2)}{6n}}
g \left[ \left( 
\frac{\pi}{n \sin \frac{\pi}{n}} 
\frac{L-l}{L} \right)^{2\Delta} - 1 
+ \mathcal{O}\left( \frac{L-l}{L} \right)^{2\Delta+1} 
\right] . 
\end{equation}
In this case, the general form of $\eta_i$ could be approximated as 
\begin{equation}
\label{eq:eta_i_large_beta}
\eta_i = \alpha_i \left( 
\frac{L-l}{L} \right)^{2\Delta} 
- g + \mathcal{O} \left( \frac{L-l}{L} \right)^{2\Delta+1} , 
\end{equation}
where $\alpha_i$ is a non-universal constant. The scaling of $\eta_i$ is then able to give the information of underlying CFT: the conformal weight $\Delta$ of first excited state and its degeneracy $g$. 
On a lattice model, one can directly diagonalize the reduced density matrix to obtain the sequenced spectrum $\{\lambda_i\}$ for both zero and finite temperature cases. 
Then the change can be calculated directly by $\eta_i = \frac{\lambda_i (\beta) - \lambda_i (0)}{\lambda_i (0)} e^{-2\pi \Delta \beta/L}$. 
With increasing the value of $\beta / L$, one should observe the validity of the analytical calculation of $\eta_i$ at lowest-order from numerical results. 
We will address this in Sec.~\ref{sec:numerics}.

\section{Numerical Simulation on the lattice model of free fermions}\label{sec:numerics} 

In this section, to check our analytical results of thermal ES, we present numerical simulations in a $(1+1)$D tight-binding model as a lattice realization of free fermionic CFT 
\begin{equation}
H = t \sum _{<i,j>} c_i^{\dagger} c_j ,
\end{equation}
where $c_i^{\dagger}$ $(c_i)$ is the creation (annihilation) operator on site $i$, and $t=1$ is the hopping amplitude between nearest sites. 
In this model, we have the lowest conformal wight $\Delta = 1/2$ and the degeneracy of the first excited state $g=4$. 
Moreover, due to the quadratic nature of the free fermions, the entanglement Hamiltonian can be written in terms of two-point correlators 
\begin{equation}
H_E = - \ln \rho_A = \ln (1-C) - \ln C , 
\end{equation}
where $C_{ij} = \langle c_i^\dagger c_j \rangle$ is the correlation matrix~\cite{Peschel_2003,Peschel_2009,Surace:2021eae}.

\subsection{Probing the CFT prediction on the thermal correction parameter $\eta_i$}

\begin{figure}[tbp]
	\centering 
	\includegraphics[width=.45\textwidth]{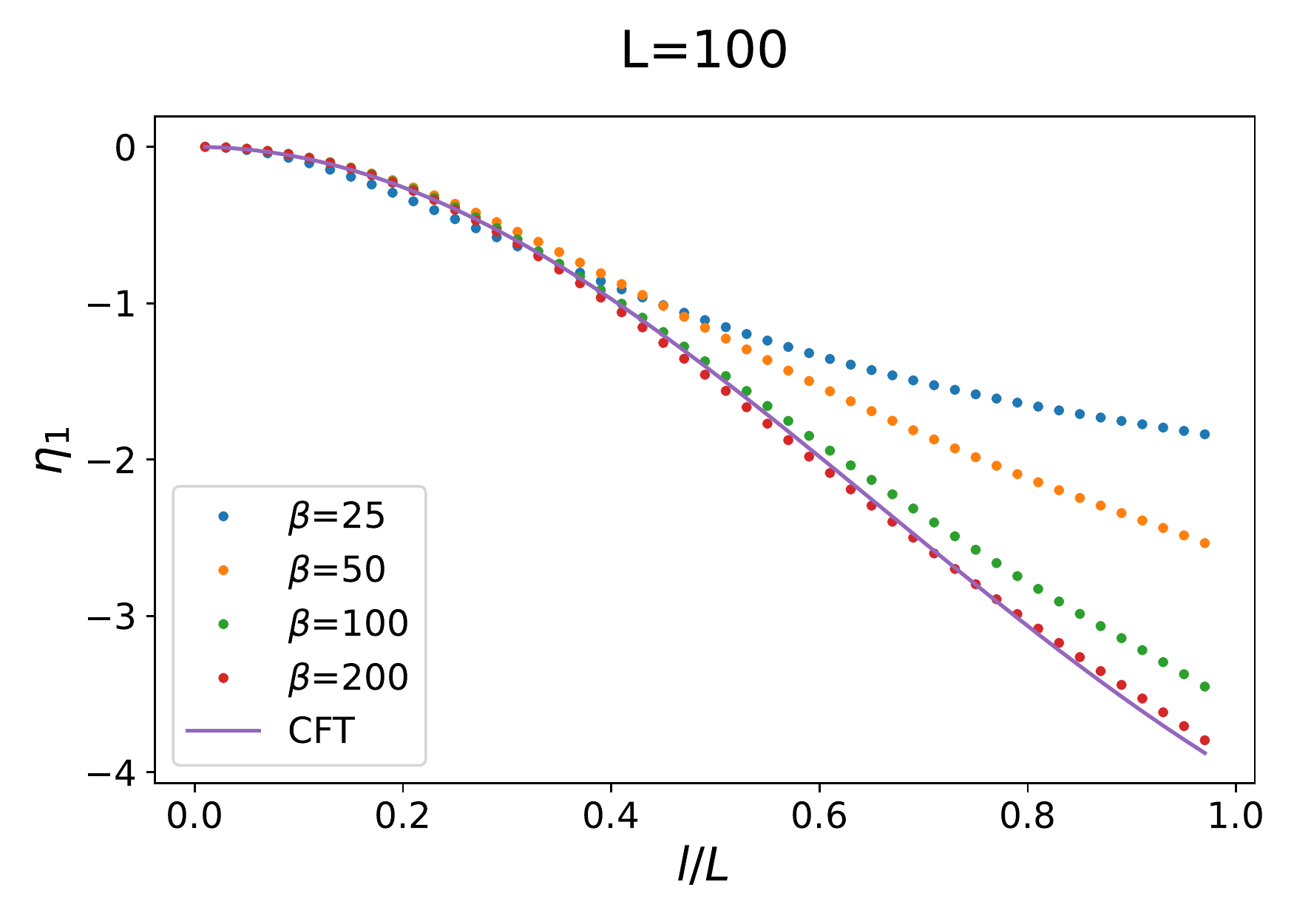}
	\hfill
	\includegraphics[width=.45\textwidth]{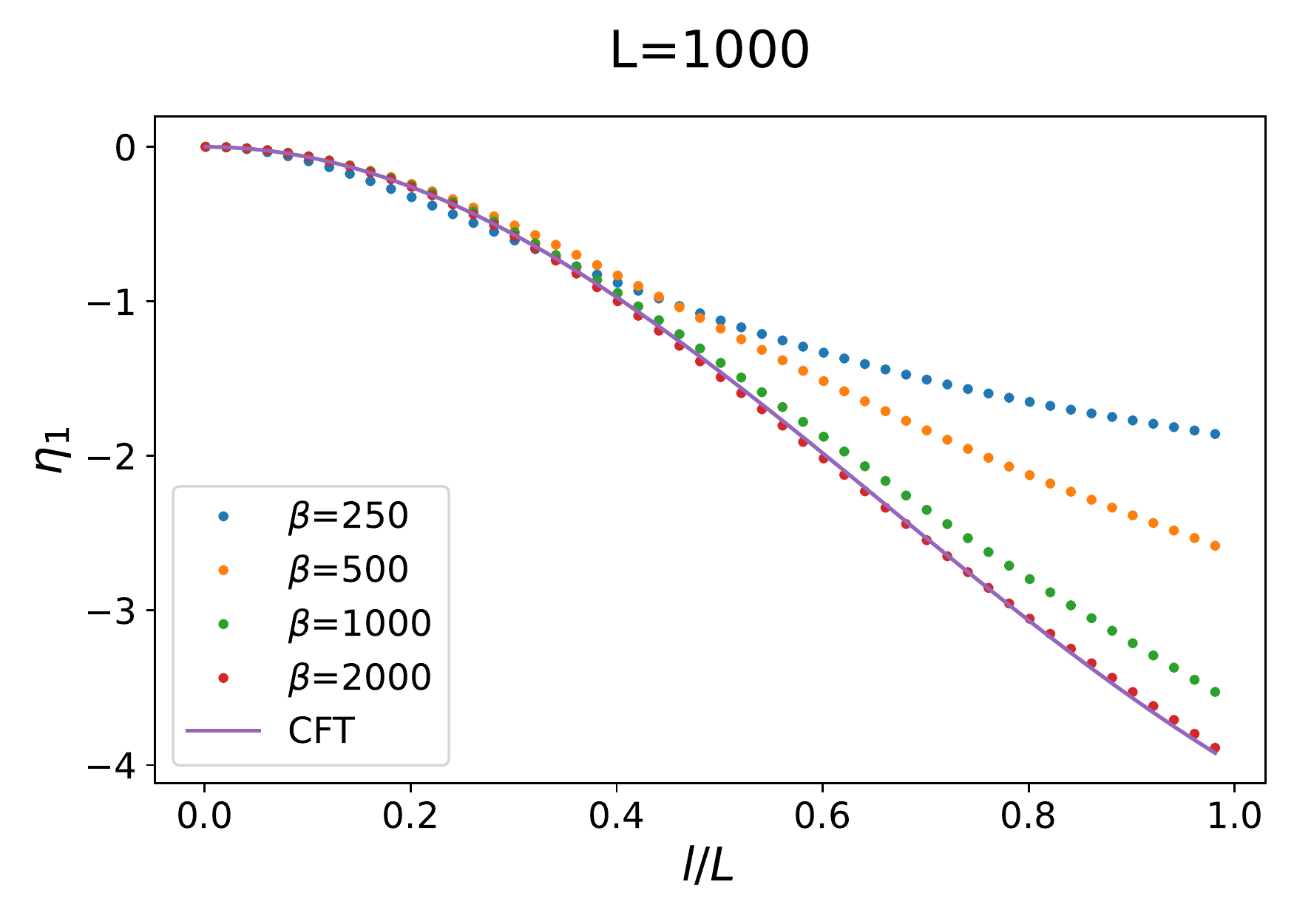}
	\\
	\includegraphics[width=.45\textwidth]{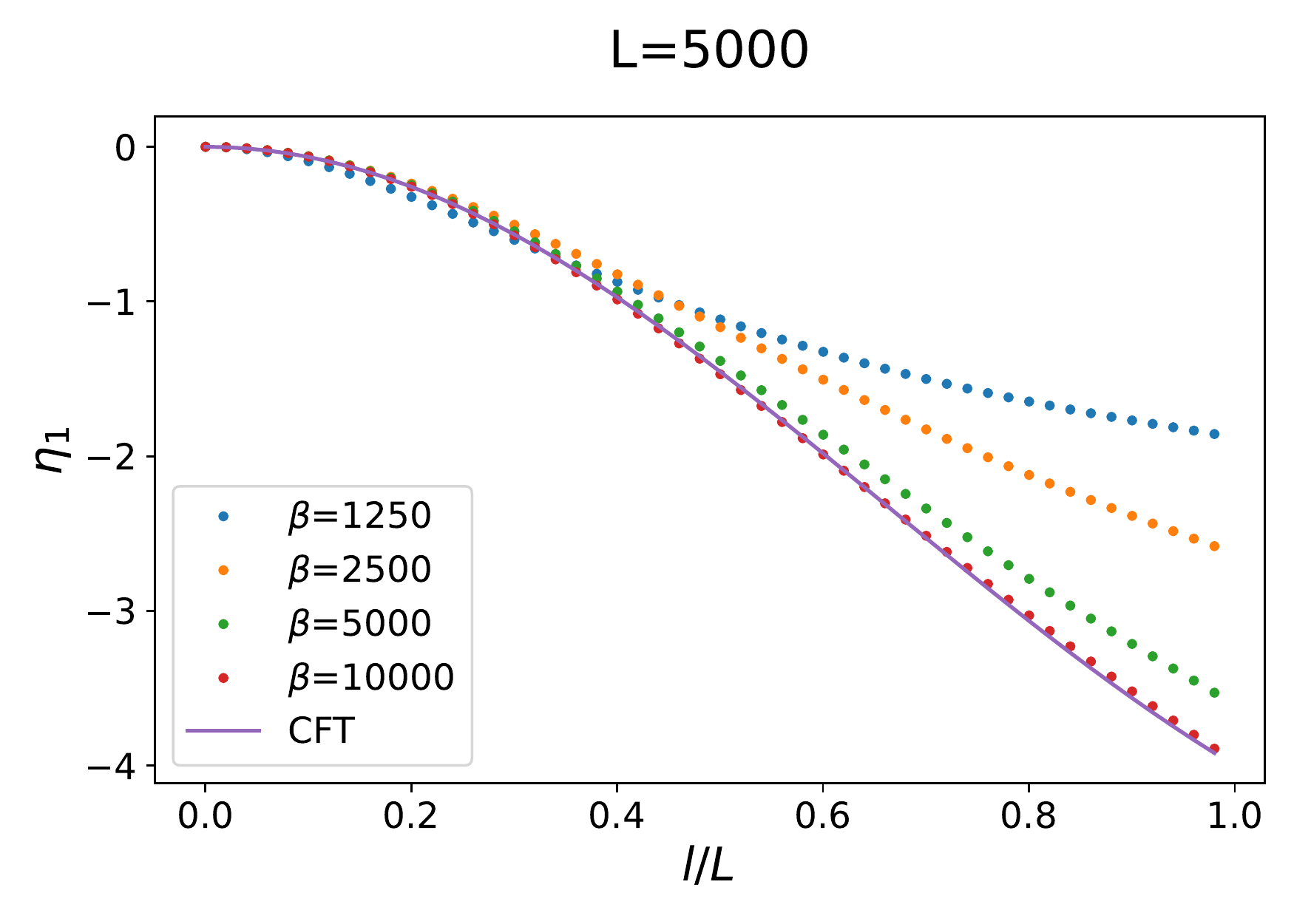}
	\hfill
	\includegraphics[width=.45\textwidth]{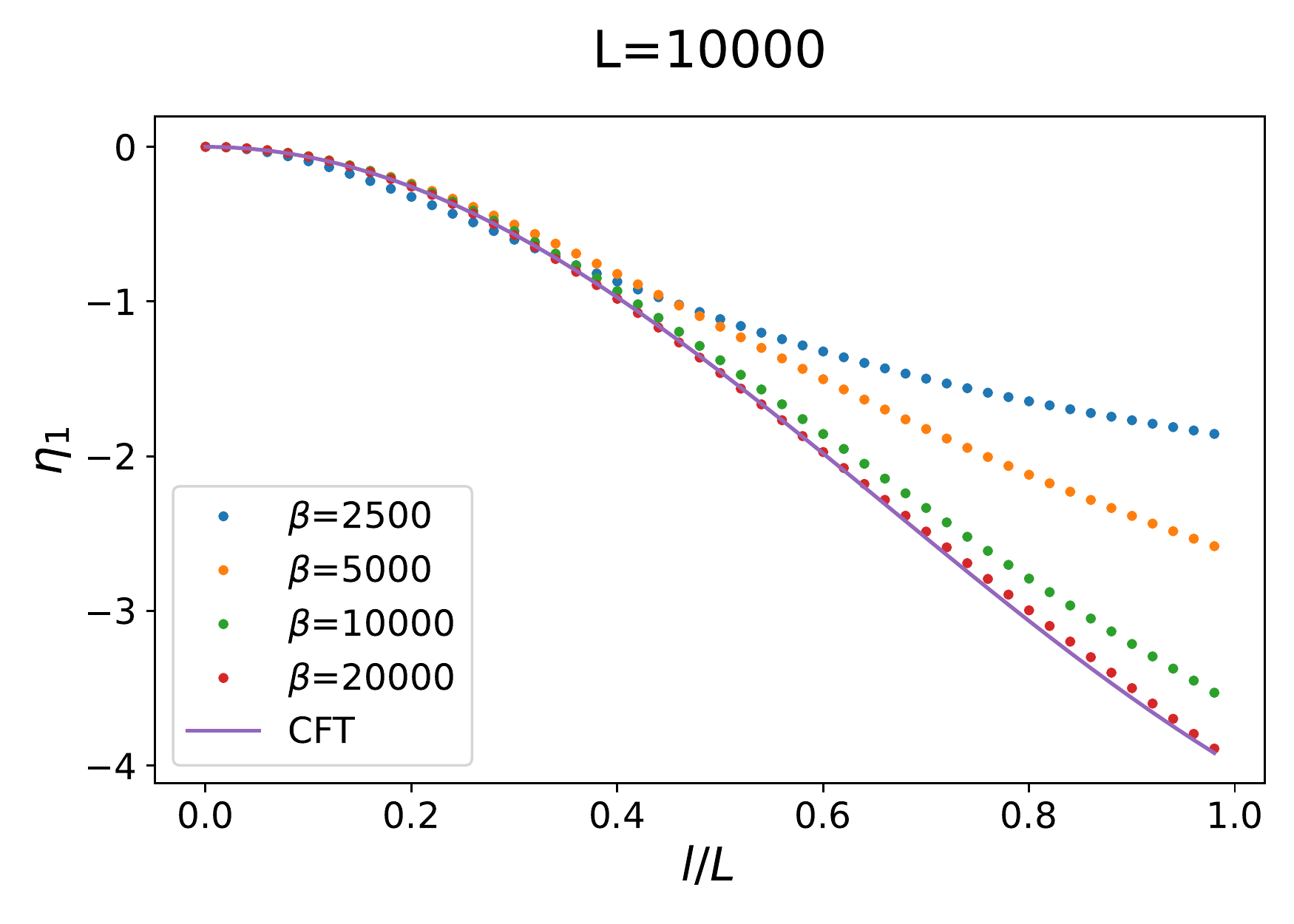}
	\caption{\label{fig:1} Numerical test of the thermal correction parameter $\eta_1$ for the total number of grids $N=100$, $1000$, $5000$ and $10000$. $\beta / L$ is fixed at $0.25$, $0.5$, $1$ and $2$ in each case. The continuous curve is the CFT prediction in  Eq. \eqref{eq:eta_1}.  }
\end{figure}

\begin{figure}[tbp]
	\centering 
	\includegraphics[width=.6\textwidth]{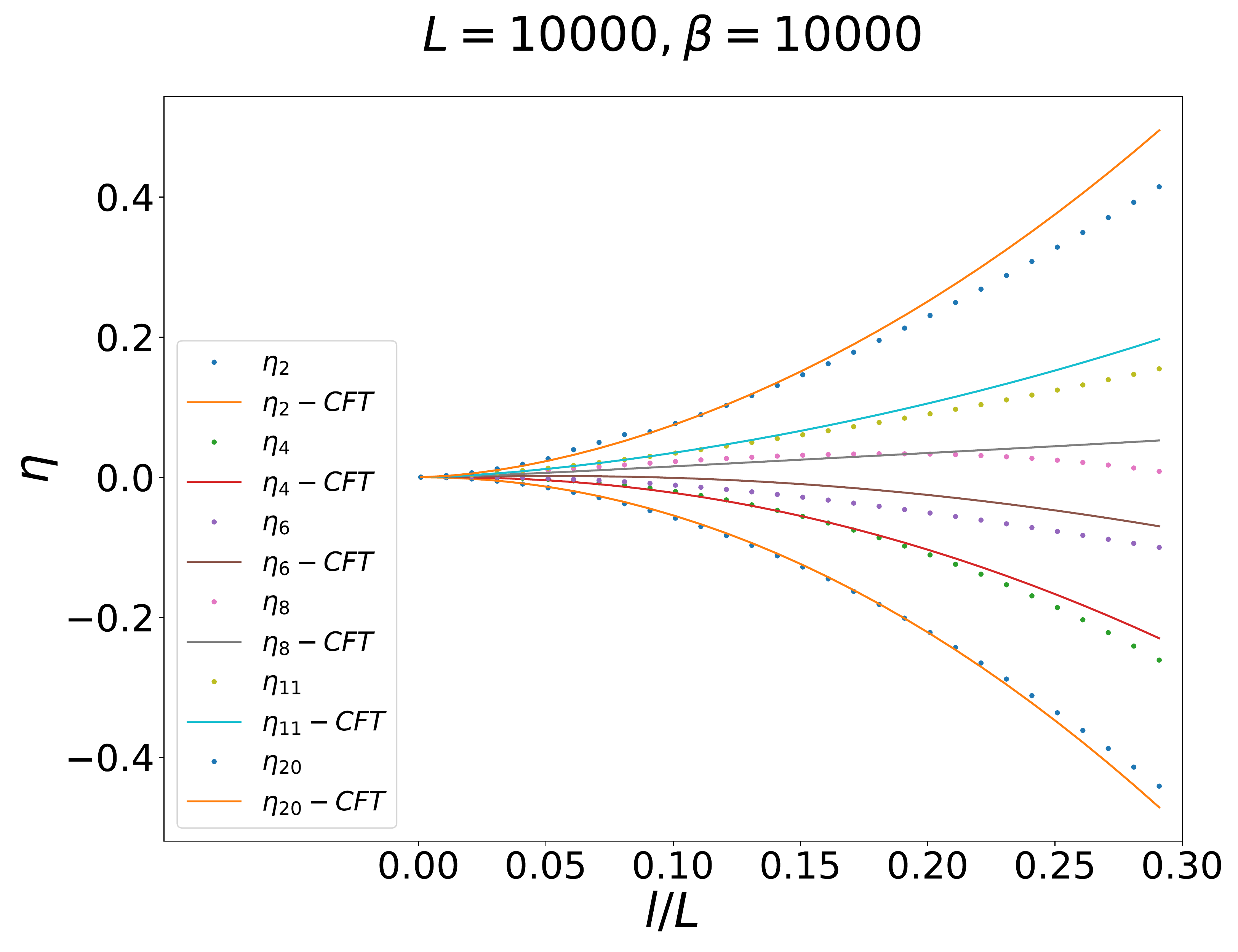}
	\caption{\label{fig:2} Numerical calculation of the thermal correction parameter from $\eta_2$ to $\eta_{20}$ for the total number of grids $N=10000$. $\beta / L$ is fixed at $1$. The continuous curve is the analytical result in \eqref{eq:eta_i}. Before the crossover ($i \leq 6$ in this figure), all eigenvalues decrease with the increase of temperature, indicating negative $\eta_i$s, while the tendency is opposite after the crossover region.}
\end{figure}

\begin{figure}[tbp]
	\centering 
	\includegraphics[width=.5\textwidth]{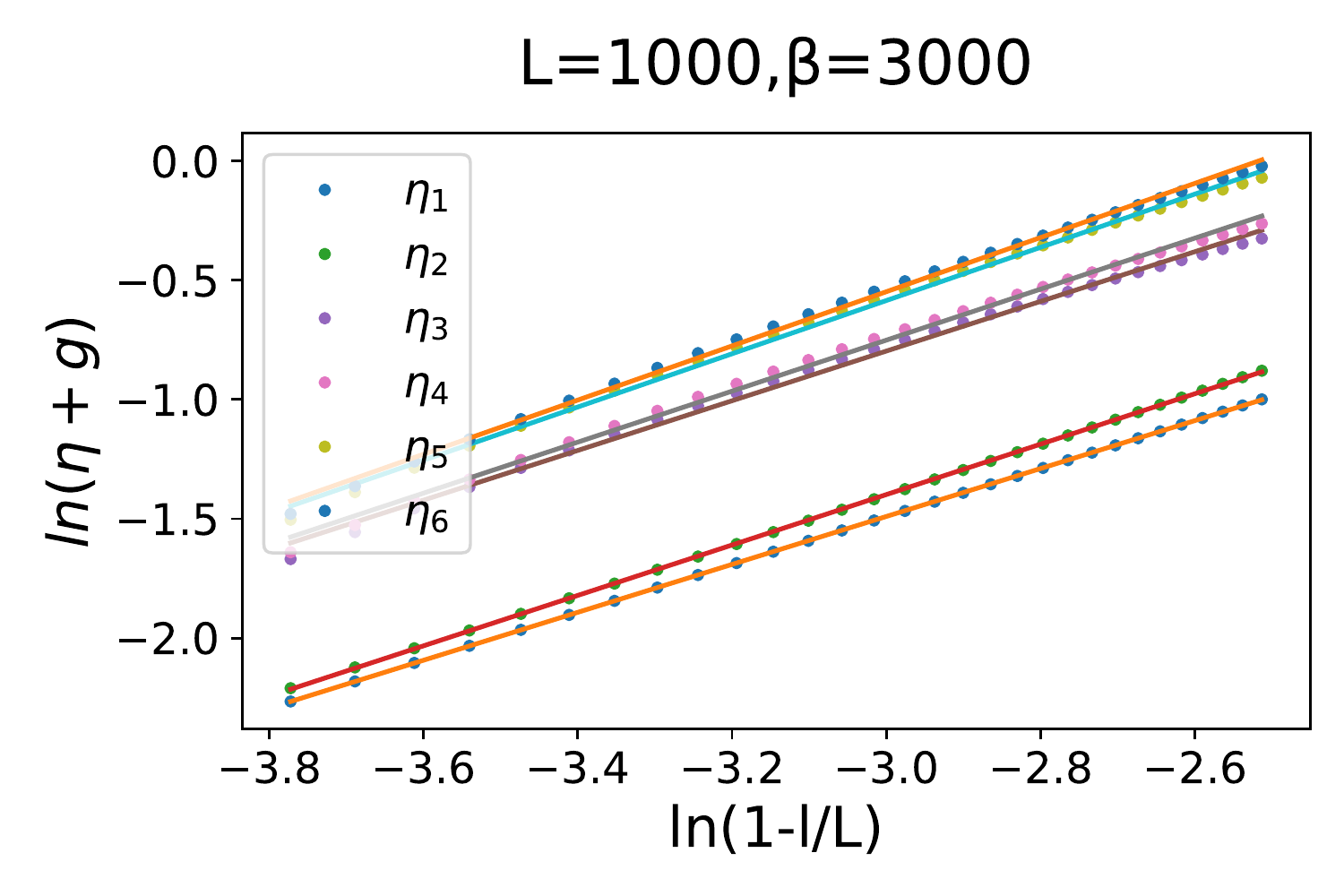}
	\caption{\label{fig:3} Numerical check of $\eta_1$ to $\eta_{6}$ for large interval limit on a $L=1000$ lattice with $\beta/L=3$. It is straightforward to get $\ln (\eta_i+g) = 2\Delta \ln(1-l/L)+c_i$ from \eqref{eq:eta_i_large_beta}. Indeed the numerical data could be fitted with a straight line with slope $2\Delta=1$.}
\end{figure}

We start with the numerical check of  the thermal parameter  $\eta_i$. 
In Eq.~\eqref{eq:eta_1}, we have obtained the analytical form of the thermal parameter $\eta_1$ for the first level of ES, which is exact for all value of the subsystem size $l$ ranging from $1$ to $L$. 
The dependence of $\eta_1$ to the subsystem size is numerically calculated in Fig.~\ref{fig:1}. 
As expected, we find that the lattice result of $\eta_1$ tends to the low-temperature CFT prediction with increasing  the ratio of $\beta/L$. 
For a value of $\beta/L = 2$, the numerical results is in good agreement with the analytical form in Eq.~\eqref{eq:eta_1}.

For solving higher levels of ES, we have performed small and large interval expansions to approach the analytical form of general $\eta_i$ in Eq.~\eqref{eq:eta_i} and~\eqref{eq:eta_i_large_beta}. 
They provide different information of the underlying finite-temperature critical theory. 
In  Fig.~\ref{fig:2}, we numerically calculate $\eta_i$ $(i \geq 2)$ with $l/L$ ranging from $0$ to $0.3$. 
The CFT result predicts a crossover of the thermal correction parameter at a certain high level of ES. 
This phenomenon is well captured by our lattice simulation. 
Before the crossover, $\eta_i$ is always negative and approaches CFT prediction with a larger $\beta$. 
Similar trend could be found after the crossover, while $\eta_i$ is positive under this circumstance. 
Near the crossover region, although the overall tendency still satisfy the CFT prediction, a deviation with the analytical result is apparent. 
This discrepancy could be traced to our definition of $\eta_i$. The thermal correction parameter should be, specifically, expressed by $\eta_i = f_{i,2} (\frac{l}{L})^2+f_{i,4} (\frac{l}{L})^4 + \cdots$, while we only keep $f_{i,2}$ term under $l/L \ll 1$. Near the turning point, $f_{i,2}$ is close to $0$, and other coefficients, like $f_{i,4}$, may not vanish, whose contribution obviously alters the values of $\eta_i$ in this region.
Meanwhile, in Fig.~\ref{fig:3}, we numerically calculate $\eta_i$ with $l/L$ changing from $0.85$ to $0.95$. 
The result is in line with our CFT derivation of Eq.~\eqref{eq:eta_i_large_beta}, which indicates the possibility of extracting universal information (the scaling dimension $\Delta$ and the degeneracy $g$ of the first excited state) of the underlying CFT.

\subsection{Thermal correction to entanglement spectrum}

\begin{figure}[tbp]
	\centering 
	\includegraphics[width=1.\textwidth]{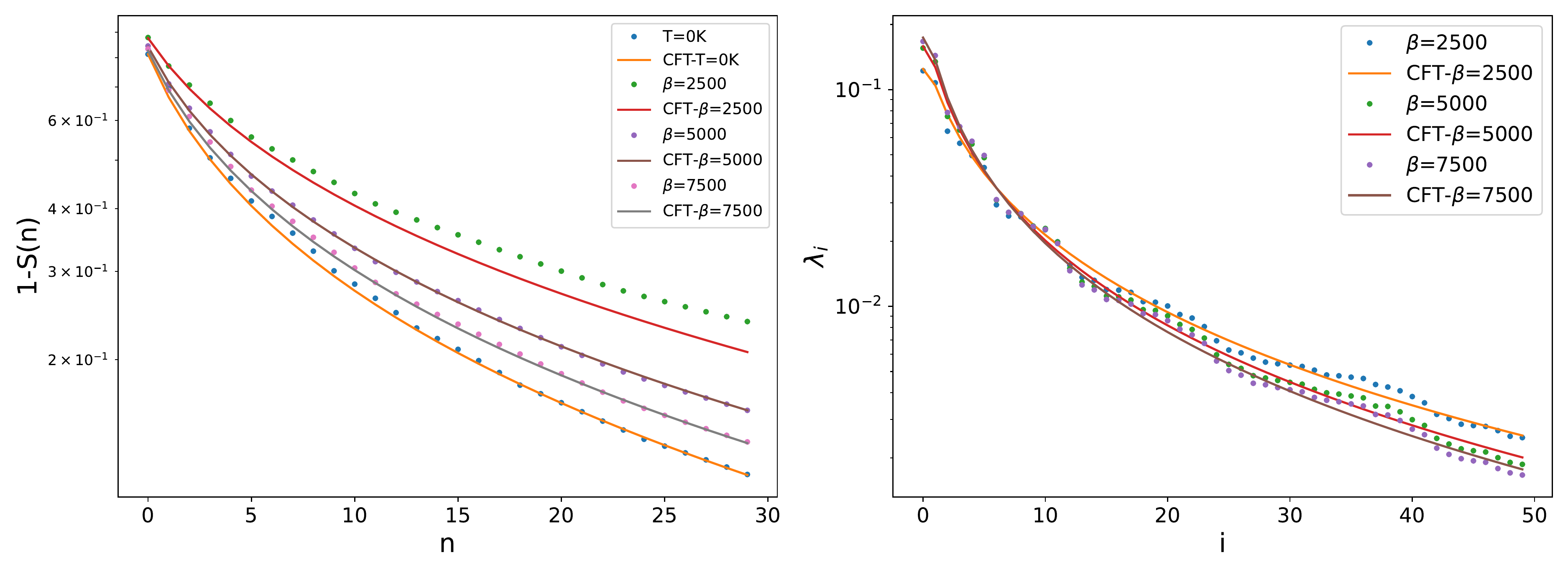}
	\caption{\label{fig:4} Thermal correction to the entanglement spectrum for $l/L=1/3$. $\beta/L$ is fixed at $0.25$, $0.5$ and $0.75$. And the entanglement spectrum for the ground state ($T=0K$) is also presented in the left diagram for contrast. Left: $1-S(n)$, where $S(n)=\sum_{i=1}^{i=n} \lambda_i$. Right: $\lambda_i$, the continuous curve is calculated by \eqref{eq:lambda_i} and \eqref{eq:eta_i}.}
\end{figure}

Here we present a direct calculation of the ES with fixed $l,L$ and $\beta$ at finite-temperature in Fig.~\ref{fig:4}.

We select 2 variables to demonstrate our prediction in Figure~\ref{fig:4}, for $l/L=1/3$, where $\{ \lambda_i \}$ are the eigenvalues of the reduced density matrix arranged from the largest one and $S(n)$ is the sum over the largest $n$ eigenvalues. 

 Additionally, the entanglement spectrum of the ground state in \eqref{eq:es_0K} is fixed by previous CFT calculation in \cite{Calabrese:2008}. When $\beta$ becomes large enough, the thermal entanglement spectrum almost conincide with the ground state entanglement spectrum, hence we choose $\beta/L = 0.25$, $0.5$ and $0.75$. It is evident our result could predict the thermal correction to the entanglement spectrum at low temperature limit. The CFT basically assume a continumm limit, while we could see the oscillation of $\lambda_i$ obtained numerically due to lattice effect. 

One key observation is the crossover, above which all eigenvalues decrease and \emph{vice versa}. 
We speculate this mode might be universal in other situation at low temperature. 
In addition, this tendency is consistent with the high temperature limit when all eigenstates are occupied by the same filling and the entanglement spectrum becomes uniform.

\section{Thermal correction to the symmetry-resolved entanglement spectrum}

In the past few years, the resolution of entanglement with $U(1)$ symmetry has attracted lots of attention from both theoretical and experimental points of view \cite{Goldstein:2017bua,Cornfeld:2018wbg,Bonsignori:2019naz,Fraenkel:2019ykl,Capizzi:2020jed,Murciano:2020lqq,Murciano:2020vgh,Parez:2020vsp,Murciano:2021djk,Capizzi:2022jpx,Ares:2022gjb,Belin:2013uta,Belin:2014mva,Zhao:2020qmn,Weisenberger:2021eby}. 
Here, we show that our approach to finite-temperature ES has a straightforward extension to the \textit{symmetry-resolved entanglement spectrum} (SRES).

For systems with an internal $U(1)$ symmetry, e.g. the particle number conservation, one allows to perform a symmetry resolution of the quantum entanglement. 
Consider the $U(1)$ symmetry is generated by a charge operator $Q$ that satisfies
$Q = Q_A \otimes \textbf{Id}_B + \textbf{Id}_A \otimes Q_B$ for a bipartition of the total system into subsystems $A$ and $B$, 
then the commutator $[\rho, Q] = 0$ implies $[\rho_A, Q_A] = 0$. 
This leads to a decomposition of the reduced density matrix into a block diagonal form $\rho_A = \oplus_q \rho_A(\rho)$ that labeled by the eigenvalue $q$ of the charge $Q$. 
The \textit{symmetry-resolved R\'enyi entropy} is then defined as  
\begin{equation}
S_n(q) = \frac{1}{1 - n} \ln \Tr 
\left[ \frac{\rho_A(q)}{p_A(q)} \right]^n 
= \frac{1}{1 - n} \ln \frac{Z_n(q)}{[Z_1(q)]^n} ,
\end{equation}
where $p_A(q) = \Tr \rho_A(q)$ is the probability that the subsystem $A$ falls into the symmetry sector $q$, and the \textit{symmetry-resolved entanglement entropy} is 
\begin{equation}
S_E(q) = \lim_{n \to 1} S_n(q) 
= - \Tr \left[ \frac{\rho_A(q)}{p_A(q)} \ln \frac{\rho_A(q)}{p_A(q)} \right] .
\end{equation}
The total EE can be written in terms of 
\begin{equation}
S_E = \sum_q p_A(q) S_E(q) - \sum_q p_A(q) \ln p_A(q) , 
\end{equation}
where the first term is called configuration entropy, and the second term is refered as fluctuation entropy that reflects the charge fluctuation between the bipartite subsystems $A$ and $B$. 
Calculating the partition function $Z_n(q)$ from a symmetry resolution in the symmetry sector $q$ is a hard task. 
It is found that the problem can be transformed to the evaluation of partition function on a replica manifold with generalized Aharonov-Bohm flux: 
$\mathcal{Z}_n(\alpha) = \Tr \left( \rho_A^n e^{i \alpha Q_A} \right) = \sum_q \Tr \left( \rho_A^n e^{i \alpha q} \right)$ \cite{Goldstein:2017bua}. 
For $(1+1)$D massless bosonic field $\phi$, the additional Aharonov-Bohm phase can be generated by inserting a vertex operator $\mathcal{V} = e^{i\frac{\alpha}{2\pi}\phi}$. 
This idea leads to an estimation of the charged moments 
\begin{equation}
\mathcal{Z}_n(\alpha) \sim l_{\text{eff}}^{\frac{c(1-n^2)}{6n}} 
l_{\text{eff}}^{-\frac{2 \left( 
		\Delta_{\mathcal{V}} + \overline{\Delta}_{\mathcal{V}} 
		\right)}{n}} ,
\end{equation}
where $l_{\text{eff}} = \frac{L}{\pi} \sin \frac{\pi l}{L} $ for cutting a single finite region with length $l$ from a ring with circumference $L$, and  $\{ \Delta_{\mathcal{V}}, \overline{\Delta}_{\mathcal{V}} \}$ is the scaling dimension of the vertex operator. 
For fermionic theories, the operator $\mathcal{V}$ can be interpreted by using bosonization relation $\psi \sim e^{i\phi}$. 
This leads to $\Delta_{\mathcal{V}} = \Delta_{\mathcal{V}} = \frac{1}{2} \left( \frac{\alpha}{2\pi} \right)^2 K$ with the Luttinger parameter $K$ that describes the interactions between fermions, which takes $K = 1$ for free fermions.

Analog to the case of total ES, for solving the thermal correction to SRES, we consider the low-temperature expansion of the charged partition function as
\begin{equation}
\mathcal{Z}_n(\alpha, \beta) 
\approx \mathcal{Z}_n(\alpha, \beta=0)
\left[ 1 + g F_n(\alpha) n e^{-2\pi \Delta \beta/L} \right] ,
\end{equation}
where 
\begin{equation}
\begin{aligned}
F_n(\alpha) & = 
\frac{\Tr \left[ 
	\left( \Tr_B |\psi \rangle \langle \psi| \right) 
	\left( \Tr_B |0 \rangle \langle 0| \right)^{n-1} 
	e^{i \alpha Q_A} \right]}
{\Tr \left[ \left( \Tr_B |0 \rangle \langle 0| \right) 
	e^{i \alpha Q_A} \right]}
- 1 \\ 
& = \frac{\sin^{2\Delta}\left( \frac{\pi l}{L} \right)}
{n^{2\Delta} \sin^{2\Delta}\left( \frac{\pi l}{nL} \right)} 
\left[ 
1 - K \left( \frac{\alpha}{\pi} \right)^2 
\sin^{2}\left( \frac{\pi l}{nL} \right) 
\right] - 1 . 
\end{aligned}
\end{equation}
which has been computed in \cite{Ghasemi:2022jxg}.

At the small interval limit $l \ll L$, we can simplify the above equation to 
\begin{equation}
F_n(\alpha)
= 
\frac{ (1-n^2) \pi^2 \Delta - 3\alpha^2 K}{3n^2} 
\left( \frac{l}{L} \right)^2 
+ \mathcal{O}\left( \frac{l}{L} \right)^4 .
\end{equation}
The charged distribution function $Q(\lambda, \alpha)$ can be derived accordingly 
\begin{equation}
\begin{aligned}
& \quad \lambda Q(\lambda, \alpha) 
= \lambda  \sum_{q} \sum_{i} \delta [\lambda - \lambda_i(q)] e^{i \alpha q} 
= \frac{1}{\pi} \lim_{\epsilon \to 0} \sum_{n=1}^\infty 
g \mathcal{Z}_n(\alpha) F_n(\alpha) (\lambda - i\epsilon)^{-n} 
\end{aligned}
\end{equation}
After some algebra, it gives 
\begin{equation}
\begin{aligned}
Q(\lambda, \alpha) \approx 
\frac{g \pi^2 l^2}{3 \lambda L^2} 
\left[ 
\theta(\lambda_1 - \lambda) I_1(2u) 
\left( \frac{A(\alpha) u}{2 B(\alpha) \ln \lambda_1} 
- \frac{2 \Delta B(\alpha) \ln \lambda_1}{u}\right) 
- \delta(\lambda_1 - \lambda) 
\Delta \lambda_1 
\right] ,
\end{aligned}
\end{equation}
where $ A(\alpha) = 1 - \frac{3 \alpha^2 K}{\pi^2} $ , 
$ B(\alpha) = \Delta - \frac{3 \alpha^2 K}{\pi^2} $ 
and $u = 2 \sqrt{\ln \lambda_1 B(\alpha) \ln \frac{\lambda_1}{\lambda}}$ . 
Analog to the case of solving the total ES, calculating the sum of first few levels allows a self-consistent check. Here, we have 
\begin{equation}
\begin{aligned}
s_\eta(M, \alpha) 
& = \int_{\lambda_M}^{\lambda_1} Q(\lambda', \alpha) d\lambda' 
= \sum_q \sum_{\lambda_i(q) \ge \lambda_M} \eta_i(q) e^{i \alpha q} \\ 
& = \frac{g \pi^2 l^2}{12 \ln^2 \lambda_1 B^2(\alpha) L^2} 
\left\{
A(\alpha) 
\left[ I_0^{-1} (M) \right]^2 
I_2\left( I_0^{-1} (M) \right) 
- 4 M \Delta \ln^2 \lambda_1 B^2(\alpha)
\right\} . 
\end{aligned}
\end{equation}
After performing a Fourier transformation from the $\alpha$ to the $q$-sector, we reach the sum of the thermal correction parameter $\eta_i(q)$ in each symmetry sector as
\begin{equation}
s_\eta(M, q) = \sum_{\lambda_i(q)\geq \lambda_M} \eta_i(q) 
= \int_{-\pi}^{\pi} \frac{d\alpha}{2\pi} e^{-i \alpha q} s_\eta(M, \alpha) 
= \int_0^\pi \frac{d\alpha}{\pi} \cos(\alpha  q) s_\eta(M, \alpha) . 
\end{equation}
As taking $M = 1$, this sum gives a single value of $\lambda_1$ as 
\begin{equation}
\eta_1 = g \left[ \left( \frac{\sin \pi l/L}{\pi l/L} \right)^{2\Delta} -  1 \right] ,
\end{equation}
which is identical to the result of $\eta_1 = \lim_{n\to\infty} g F_n(\alpha)$. Moreover, it is easy to check that by taking $\alpha = 0$, the above calculations are reduced to the case of total ES as discussed in previous section.

\section{Discussion}
\label{sec:4}

In this paper, we have derived the thermal correction to the entanglement spectrum for two dimensional CFTs.
With the small interval and low-temperature expansion, the thermal correction of order $o(e^{-2\pi \Delta \beta/L})$ has been analytically calculated for each eigenvalue. 
Interestingly, at the large interval limit, the correction displays some scaling behavior depending on the scaling dimension $\Delta$ and the degeneracy $g$ of the first excited state. 
This allows to extract the universal information of the underlying CFT. 
Moreover, in all cases, we find a clear crossover changing pattern inside the entanglement spectrum, which encodes how thermal effect reduces the underlying entanglement structure within critical systems. 
All of the above analytical predictions have been verified in our numerical simulation of a lattice model that realizes a free fermionic CFT.

For zero-temperature cases~\cite{Calabrese:2008}, the conformal symmetry in two dimensional theories is found to be strong enough to determine the entire entanglement spectrum by merely the central charge $c$ of the underlying CFT.
As discussed in the main text, with considering the finite-temperature correction, the information of excited states comes into the entanglement spectrum. 
At low-temperature limit, the correction is dominated by the first excited state and depends on its scaling dimension $\Delta$ and degeneracy $g$. 
This leads to the possibility of extracting universal information of underlying CFT via the thermal entanglement  spectrum.

We expect more practical applications for the present result on thermal correction of the entanglement spectrum. 
One possible direction is to investigate the thermalization process of the system from the thermal correction to the entanglement spectrum. 
Since the expectation value of any operator $\hat{O}$ defined on the subsystem $A$ depends linearly on $\rho_A$ as $\langle \hat{O} \rangle = \Tr(\rho_A \hat{O})$, it is then natural to consider an estimation of these expectation values from the spectral information of reduced density matrix. 
For example, it is straightforward to know that $ (\langle \hat{O}(\beta) \rangle -\langle \hat{O}(T=0) \rangle) \propto \Tr \rho_A(\beta) - \Tr \rho_A(T=0)  \propto e^{-2\pi \Delta \beta/L}$. 
A careful study would provide more detailed information about thermalizing critical systems from the perspective of quantum entanglement. 

We end up with some future directions. Within current work, we focus on two-dimensional case, while similar thermal correction might be able to get for some special cases in higher dimensional spheres $\mathbb{S}^{d-1}$ \cite{Herzog:2014fra,Herzog:2014tfa}. Secondly, we consider low temperature limit, where the ground state and the first excited states dominates the behaviour of entanglement spectrum. For a generic low-energy excited state, the $n$-th order R\'enyi entropy could be computed through a $2n$-point correlation function \cite{Alcaraz:2011tn}. Additionally, for more general temperature, the R\'enyi entropy for free fermions or free bosons at $\mathbb{S}^1$, which involves either Jacobi elliptic theta functions or Riemann-Siegel functions, was computed through the correlation function of twistor operators in high genus surface \cite{Azeyanagi:2007bj,Herzog:2013py,Datta:2013hba,Mukhi:2017rex}. It is meaningful to calculate the thermal entanglement spectrum within corresponding cases based on previous work. Thirdly, recent investigation motivated more characterization of entanglement structure for critical systems, such as entanglement negativity \cite{Calabrese:2012ew,Calabrese:2012nk}. It is interesting to know how underlying information emerges in thermal negativity spectrum comparing with the ground state negativity spectrum \cite{Ruggiero:2016yjt,Shapourian:2019xfi}. However, only limited results have been achieved for finite temperature R\'enyi negativity so far \cite{Murciano:2021djk,Calabrese:2014yza,Shapourian:2018lsz,Wu:2019qvm}. Finally, it was pointed out that the ground state entanglement spectrum could reproduce more information, such as operator content \cite{Lauchli:2013jga} and Affleck-Ludwig entropy \cite{Alba:2017bgn}, besides the central charge of the underlying CFT. Similar study might unveil more details in the thermal entanglement spectrum, which we leave as future work.

\section*{Acknowledgements}
This work was supported by ``Pioneer" and ``Leading Goose" R\&D Program of Zhejiang (2022SDXHDX0005), the Key R\&D Program of Zhejiang Province (2021C01002) and the foundation from Westlake University.
We thank Westlake University HPC Center for computation support.

\appendix

\bibliography{Thermal_Correction_ES}

\end{document}